# Nonlinear dust-acoustic solitary waves and shocks in dusty plasmas with a pair of trapped ions

Nirab C. Adhikary[1, #], Amar P. Misra[2], Manoj K. Deka[3] and Apul N. Dev[4]

[1]*Physical Sciences Division, Institute of Advanced Study in Science and Technology, Vigyan Path, Paschim Boragaon, Garchuk, Guwahati-781035, Assam, India*
[2]*Department of Mathematics, Siksha Bhavana, Visva-Bharati University, Santiniketan-731 235, India.*
[3]*Department of Applied Sciences, Institute of Science and Technology, Gauhati University, Guwahati-781014, Assam, India*
[4]*Center for Applied Mathematics, Siksha 'O' Anusandhan University, Khandagiri, Bhubaneswar-751030, Odisha, India.*

#corresponding author's e-mail: nirab_iasst@yahoo.co.in

**Abstract:**

The propagation characteristics of small-amplitude dust-acoustic (DA) solitary waves (SWs) and shocks are studied in an unmagnetized dusty plasma with a pair of trapped positive and negative ions. Using the standard reductive perturbation technique with two different scaling of stretched coordinates, the evolution equations for DA SWs and shocks are derived in the forms of complex Korteweg-de Vries (KdV) and complex Burgers' equations. The effects of dust charge variation, the dust thermal pressure, and the ratios of the positive to negative ion number densities as well as the free to trapped ion temperatures on the profiles of SWs and shocks are analysed and discussed.

**I. Introduction:**

Ever since the prediction of the imperative roles of trapped plasma particles on the propagation characteristics of solitary waves by Schamel in 1972 [1-3], there has been a number of works in the literature reported on the study of plasma wave spectra in presence of trapped plasma ions and electrons [4-14]. Such trapped particles significantly alter the nonlinearity of a plasma system thereby bringing momentous modification in the dynamical evolution of solitary and shock wave phenomena. It has been well established that charged particle trapping is inevitable in space and laboratory plasmas [15,16] and there has been a number of works both on theoretical [4-14] and experimental aspects of trapped particles states [17-19]. In a recent report, Dev *et al.* [6] discussed the behaviours of both compressive and rarefective shock waves by deriving a three-dimensional (3D) Burgers' equation with Maxwellian electrons and non-thermal ions. They pointed out that the evolution of only rarefactive shock waves are possible due to the effects of trapped ions and isothermal electrons. On the other hand, El-Hanbaly *et al* [20] studied the properties of both solitary waves (SWs) and shocks in dissipative plasmas in presence of trapped ions. In an aother work, Rahman and Haider [21] studied the effect of trapped negative ions on the propagation of dust- ion acoustic (DIA) SWs in the description of a modified Korteweg-de Vries (KdV) equation. They concluded that the trapped state of negative ions give rise to a stronger nonlinearity i.e., the evolution of solitary structures with larger amplitude, smaller width, and higher propagation velocity. Recently, Deka [7] discussed the effect of trapped electrons and non-thermal ions on dust-acoustic (DA) SWs (DASWs) and it was shown that only compressive DASWs exist due to possible interplay between nonthermality and trapped state of the plasma. Solitary wave phenomena in pair-ion plasmas have also been discussed extensively in different contexts. To mention few, Misra and Adhikary [22] discussed the propagation of SWs in a pair-ion plasma and they found that such plasmas support only the rarefactive solitons in presence of positively charged dust grains. Also, Adak *et al* [23] studied the propagation of shocks in a collisional pair-ion plasma. Furthermore, Misra and Barman [24] discussed the effects of obliqueness and magnetic field on the propagation of DIA SWs in magnetized pair-ion plasmas and concluded that the transition of solitary wave from rarefactive to compressive depends on the mass ratio of negative to positive ions.

Recently, Misra [25] studied the evolution of DASWs in a magnetized dusty pair-ion plasma with vortex-like distribution of ions. It was shown that the nonlinear coefficient of the KdV equation becomes complex due to the trapped states of both positive and negative ions. He also showed that apart from the external magnetic field and obliqueness of propagation, the trapped ion temperatures, the dust as well as the ion thermal pressure and the density of positive ions have profound effects on the amplitude and width of the solitons. Furthermore, the nonlinear properties of ion-acoustic shock waves have been studied by Abdelwahed *et al*. [26] taking into account the effects of time fractional parameter. The results were shown to be important in the D- and F-regions of the Earth's ionosphere. In an another work, Abdelwahed *et al. [27]* investigated the propagation of rogue waves in superthermal plasmas with a pair of ions. Different characteristics of the rogue waves so generated have been studied by the effects of the number densities as well as the superthermality of ions.

Apart from these investigations, a few works on SWs in dusty pair-ion plasmas including the effects of degeneracy, non-thermality as well as the supra-thermality of plasma parameters have also been carried out [28-30]. On the other hand, a few attempts have been made to solve the complex KdV and complex Burgers' equations throughout the research fraternity from different branches of physics as well as mathematics considering different techniques in different physical situations [31-35].

In this work, we present a theoretical investigation on the characteristics of small-amplitude DASWs and shocks in an unmagnetized dusty pair-ion plasma with the effects of trapped particle distributions of ions and dust charge fluctuations. Starting from a set of fluid equations for charged dust grains and Vlasov equations for trapped ions we derive a 3D complex KdV and complex Burgers' equations for the evolution of DASWs and shocks in dusty pair-ion plasmas. . We also study the effects of the dust charge variation, the dust thermal pressure, and the ratio of the positive to negative ion number densities as well as the ratio of free to trapped ion temperatures on the profiles of SWs and shocks.

**II. Basic equations:**

We consider the nonlinear propagation of small-amplitude DASWs and shocks in a 3D dusty electron free pair-ion plasma with the effects of vortex-like distributions ions and dust charge fluctuations. The collisions between all the particles are considered to be negligible compared to the dust plasma period. Furthermore, in dusty pair-ion plasmas the ratio of electric charge to mass of dust particles assumed to be much smaller than those of positive and negative ions. The basic equations for the dynamics of charged dust grains are

$$\frac{\partial n_d}{\partial t} + \nabla \cdot (n_d v_d) = 0, \quad \text{------ (1)}$$

$$\frac{\partial v_d}{\partial t} + (v_d \cdot \nabla) v_d = \frac{q_d}{m_d} E - \frac{\nabla P_d}{m_d n_d}, \quad \text{------ (2)}$$

$$\frac{\partial^2 \psi}{\partial x^2} = -4\pi e (n_p - n_n + \alpha Z_d n_d), \quad \text{------ (3)}$$

where $n_d$ is the dust number density normalized by its equilibrium value $n_{d0}$, $v_d$ is the dust fluid speed normalized by the DA speed $c_d = \sqrt{Z_d \kappa_B T_p / m_d} = \omega_{pd} \lambda_D$, with $\omega_{pd} = \sqrt{4\pi n_{d0} Z_d^2 e^2 / m_d}$ and $\lambda_D = \sqrt{\kappa_B T_p / 4\pi n_{d0} Z_d e^2}$ denoting, respectively, the dust plasma frequency and the plasma Debye length. Here, $\kappa_B$ is the Boltzmann constant and $T_p$ is the thermodynamic temperature of free positive ions. Also, $q_d = \alpha Z_d e$ is the charge on the dust grain, where $\alpha = \pm 1$ stands for positively or negatively charged dust grains and $Z_d$ is the dust charge state, i.e., the collection of positive or negative ions by dust grains. Also $E = -\nabla \varphi$ is the electric filed with $\varphi$ denoting the electrostatic scalar potential and $P_d$ is the dust thermal pressure given by the adiabatic equation of state $P_d / P_0 = (n_d / n_{d0})^\gamma$. Here, $\gamma = (2+D)/D = 5/3$ is the adiabatic index for 3D fluid flow, $P_0 = n_{d0} \kappa_B T_d$ is the equilibrium dust pressure and $T_d$ is the thermodynamic temperature of dust grains. In dimensionless variables, Eqs. (1) - (3) can be recast as

$$\frac{\partial N_d}{\partial T} + \nabla \cdot (N_d V_d) = 0 \quad \text{----- (4)}$$

$$\frac{\partial V_d}{\partial T} + V_d \cdot \nabla V_d + \alpha Z_d \nabla \phi = -\frac{5}{3} \sigma_d N_d^{-1/3} \nabla N_d \quad \text{----- (5)}$$

$$\nabla^2 \phi = (\mu_n n_n - \mu_p n_p - \alpha Z_d N_d) \quad \text{----- (6)}$$

Where $\sigma_d = T_d / T_p$, $\mu_p = n_{p0} / Z_{d0} n_{d0}$ and $\mu_n = n_{n0} / Z_{d0} n_{d0}$ are the equilibrium number density ratios with $n_{p0}$ and $n_{n0}$ denoting the number density for trapped positive and negative ions which satisfy the charge neutrality condition at equilibrium, $\mu_n - \mu_p = \alpha$.

In the small-amplitude limit with $|\sigma \varphi| < 1$, the expressions for the number densities $n_p \& n_n$ of trapped positive and negative ions can be obtained as [1, 25],

$$n_p = 1 + (-\phi) - b_p (-\phi)^{\frac{3}{2}} + \frac{1}{2}(-\phi)^2 - \ldots, \quad b_p = \frac{4(1-\sigma_p)}{3\sqrt{\pi}} \quad \text{----- (7)}$$

$$n_n = 1 + \sigma \phi - b_n (\sigma \phi)^{\frac{3}{2}} + \frac{1}{2}(\sigma \phi)^2 - \ldots, \quad b_n = \frac{4(1-\sigma_n)}{3\sqrt{\pi}} \quad \text{----- (8)}$$

where $\sigma_p = T_{efp} / T_p$ and $\sigma_n = T_{efn} / T_n$ with $T_n$, $T_{efp}$ and $T_{efn}$ are the is the trapped negative ion, free positive ion and free negative ion temperatures respectively. We assume that the charging of the dust gains arise due to the flow of positive ions and negative ions into the grain surface. Here, we note that since positive ions are the lighter species, dust grains can be positively charged. On the other hand it can have negative charges due to higher negative ion concentration. Now, the charge on a spherical (isolated) dust grain is given by $q_d = a\varphi_s$, where $\varphi_s$ is the dust grain floating (surface) potential relative to the plasma potential and '$a$' represents the size of dust grain. Since the potential $\varphi_s$ can be positive or negative, depending upon the conditions, the expressions for the positive and negative ion currents will be different for $\varphi_s > 0$ and $\varphi_s < 0$. For $\varphi_s > 0$, the negative ions are accelerated towards the dust gains and positive ions are repelled. However, an opposite trend occurs for $\varphi_s < 0$, *i.e.*, the positive ions are accelerated towards the dust grain and negative ions are repelled. We assume that the repelled species can be treated using the Boltzmann distribution and the attracted species follow the usual OML theory. Thus, for $\varphi_s < 0$ *i.e.*, when the dust grains are positively charged, the expression for the ion currents are

$$I_p = 4\pi e a^2 \left(\frac{\kappa_B T_p}{2\pi m_p}\right)^{1/2} n_p(\phi, \sigma_p) \left(1 - \frac{e\phi_s}{\kappa_B T_p}\right) \quad \text{-----(9)}$$

$$I_n = -4\pi e a^2 \left(\frac{\kappa_B T_n}{2\pi m_n}\right)^{1/2} n_n(\phi, \sigma_n) \left(\frac{e\phi_s}{\kappa_B T_n}\right) \quad \text{-----(10)}$$

On the other hand, for $\phi_s > 0$ *i.e.*, when the dust grains are negatively charged, the expression for the ion currents are

$$I_p = 4\pi e a^2 \left(\frac{\kappa_B T_p}{2\pi m_p}\right)^{1/2} n_p(\phi, \sigma_p) \left(-\frac{e\phi_s}{\kappa_B T_p}\right) \quad \text{-----(11)}$$

$$I_n = -4\pi e a^2 \left(\frac{\kappa_B T_n}{2\pi m_n}\right)^{1/2} n_n(\phi, \sigma_n) \left(1 + \frac{e\phi_s}{\kappa_B T_n}\right) \quad \text{-----(12)}$$

The dust charge fluctuation is given by

$$\left(\frac{\partial}{\partial t} + v_d.\nabla\right) q_d = I_p + I_n. \quad \text{-----(13)}$$

Thus, in dimensionless variables the Eq. (13) reduces for positively and negatively charged dust grains, respectively, are

$$\left(\frac{\partial}{\partial T} + v_d.\nabla)\right) Z_d = n_p(\phi, \sigma_p)(1 - Z_d \beta_p) - n_n(\phi, \sigma_n)\exp(Z_d \beta_n) \quad \text{-----(14)}$$

$$\left(\frac{\partial}{\partial t} + v_d.\nabla)\right) Z_d = n_p(\phi, \sigma_p)\exp(Z_d \beta_p) - n_n(\phi, \sigma_n)(1 - Z_d \beta_n) \quad \text{-----(15)}$$

where $\beta_p = e^2 / \kappa_B a T_p$ and $\beta_n = e^2 / \kappa_B a T_n$

At equilibrium, the charge current balance reduces to $I_{p0} + I_{n0} = 0$. In Secs. IV and V, we use two different scaling for the stretched coordinates when the dust charging rate is smaller than and comparable to the dust plasma period.

### III. Derivation of KdV equation

In order to derive the evolution equation for the nonlinear propagation of small-amplitude DA waves, we use the standard reductive perturbation technique in which the independent variables $\xi$ and $\tau$ are stretched as

$$\xi = \varepsilon^{1/4}\left(I_x x + I_y y + I_z z - MT\right) \text{ and } \tau = \varepsilon^{3/4}T, \quad \text{----- (16)}$$

where $M$ is the Mach Number (normalized by $c_d$) and $\varepsilon$ is a small nonzero constant measuring the weakness of the perturbation. The dependent variables $N_p$, $V_p$, $N_n$, $V_n$ and $\phi$ can be expanded in power series of $\varepsilon$ as

$$N_d = 1 + \varepsilon N_d^{(1)} + \varepsilon^{3/2} N_d^{(2)} + ....,$$
$$V_{dx} = \varepsilon V_d^{(1)} + \varepsilon^{3/2} V_d^{(2)} + ....,$$
$$\phi = \varepsilon \phi^{(1)} + \varepsilon^{3/2}\phi^{(2)} + \varepsilon^2 \phi^{(3)}....,\quad \text{----- (17)}$$
$$Z_d = 1 + \varepsilon Z_d^{(1)} + \varepsilon^{3/2} Z_d^{(2)} + ....,$$
$$V_{dx,y} = \varepsilon^{5/4} V_d^{(1)} + \varepsilon^{7/4} V_d^{(2)} + .....$$

Now, substituting Eqs. (16) and (17) into Eqs. (4) – (6) and (13, and equating lowest powers of $\varepsilon$, we obtain the following first order quantities

$$N_d^{(1)} = \frac{I_x}{M} V_{dx}^{(1)} = \left(\sigma\mu_n + \mu_p + \frac{(1+\sigma)}{(\beta_p + \beta_n)}\right)\phi^{(1)}, \quad Z_d^{(1)} = -\alpha\frac{(1+\sigma)}{(\beta_p + \beta_n)}\phi^{(1)} \quad \text{----- (18)}$$

and the dispersion relation for the nonlinear wave speed given by

$$M = I_x \left[\frac{5\sigma_d}{3} + \left(\frac{(\beta_p + \beta_n)}{(\beta_p + \beta_n)(\sigma\mu_n + \mu_p) + (1+\sigma)}\right)\right]^{1/2} \quad \text{----- (19)}$$

Equation (19) describes the phase velocity of the nonlinear DA waves in dusty pair-ion plasmas, which typically depends on the dust temperature, the temperatures of positive and negative ions as well as the particle number densities. Inspecting the terms in the square brackets, we find that the second term in the parentheses is smaller than the unity. Also, the first term, proportional to $\sigma_d$, is <1 for $\sigma_d$ <0.6. Thus, for $\sigma_d \ll 1$ and $I_x<1$, Eq. (19) gives $M<1$. This implies that the propagation of DA waves in dusty pair-ion plasmas is always subsonic. The profiles of the phase velocity $M$ are shown in Fig.1 for different values of $\mu_n$. It is found that the values of $M$ is <1 and it decreases with increasing values of σ, however, increases with decreasing values of $\mu_n/\mu_p$.

Equating the coefficients of the next higher order of $\varepsilon$ we obtain from Eqs. (4) – (6) the following equations

$$M\frac{\partial N_d^{(2)}}{\partial \xi} = \frac{\partial N_d^{(1)}}{\partial \tau} + \sum_{j=x,y,z} I_j \frac{\partial}{\partial \xi}V_{dj}^{(2)} \quad \text{----- (20)}$$

$$M\sum_{j=x,y,z}\frac{\partial V_{dj}^{(2)}}{\partial \xi} = \frac{\partial V_{dx}^{(1)}}{\partial \tau} + \sum_{j=x,y,z} I_j\left(\alpha\frac{\partial \varphi^{(2)}}{\partial \xi} + \frac{5\sigma_d}{3}\frac{\partial N_d^{(2)}}{\partial \xi}\right) \quad \text{----- (21)}$$

$$I_x^2 \frac{\partial^2 \phi^{(1)}}{\partial \xi^2} = \left(\sigma\mu_n + \mu_p\right)\phi^{(2)} - b_n\mu_n(\sigma)^{3/2}\left(\phi^{(1)}\right)^{3/2} + b_p\mu_p\left(-\phi^{(2)}\right)^{3/2} - \alpha\left(N_d^{(2)} + Z_d^{(2)}\right) \quad \text{----- (22)}$$

$$\frac{\partial Z_d^{(2)}}{\partial \xi} = -\frac{\alpha(1+\sigma)}{(\beta_p+\beta_n)}\frac{\partial \phi^{(2)}}{\partial \xi} - \frac{\alpha b_p}{(\beta_p+\beta_n)}\left(-\phi^{(1)}\right)^{\frac{1}{2}}\frac{\partial \phi^{(1)}}{\partial \xi} + \frac{\alpha b_n}{(\beta_p+\beta_n)}(\sigma)^{\frac{3}{2}}\left(\phi^{(1)}\right)^{\frac{1}{2}}\frac{\partial \phi^{(1)}}{\partial \xi} \quad \text{-----(23)}$$

Note that in Eq. (23), the space-time evolution part appears corresponding to the second order perturbation of the dust charge state. It follows that in the evolution equation for the first order perturbations, the dynamical part of the dust charge number will not contribute and so no dissipative term in the evolution equation. Eliminating the second order perturbations $V_j^{(2)}$ and $N_d^{(2)}$ from Eqs. (20) - (23) and using the results of Eqs. (18) and (19) we obtain the following complex KdV equation for positively and negatively charged dust grains $(\alpha = \pm 1)$.

$$\frac{\partial \phi^{(1)}}{\partial \tau} + \left\{A_n\left(\phi^{(1)}\right)^{\frac{1}{2}} + A_p\left(-\phi^{(1)}\right)^{\frac{1}{2}}\right\}\frac{\partial \phi^{(1)}}{\partial \xi} + B\frac{\partial^3 \phi^{(1)}}{\partial \xi^3} = 0 \quad \text{-----(24)}$$

$$A_p = -\frac{\left[\{b_{1p}/(\beta_p+\beta_n)\} + \frac{3}{2}b_p\mu_p\right](\beta_p+\beta_n)(3M^2 - I_x^2 5\sigma_d)}{6M\{(\sigma\mu_n+\mu_p)(\beta_p+\beta_n)+(1+\sigma)\}} \quad \text{-----(25)}$$

$$A_n = \frac{\left[\{b_{1n}/(\beta_p+\beta_n)\} + \frac{3}{2}b_{1n}\mu_n\right](\beta_p+\beta_n)(3M^2 - I_x^2 5\sigma_d)}{6M\{(\sigma\mu_n+\mu_p)(\beta_p+\beta_n)+(1+\sigma)\}}(\sigma)^{\frac{3}{2}} \quad \text{-----(26)}$$

$$B = \frac{(\beta_p+\beta_n)(3M^2 - I_x^2 5\sigma_d)}{6M\{(\sigma\mu_n+\mu_p)(\beta_p+\beta_n)+(1+\sigma)\}} \quad \text{-----(27)}$$

Equation (24) can be rewritten as

$$\frac{\partial \phi^{(1)}}{\partial \tau} + A\left(\phi^{(1)}\right)^{\frac{1}{2}}\frac{\partial \phi^{(1)}}{\partial \xi} + B\frac{\partial^3 \phi^{(1)}}{\partial \xi^3} = 0 \quad \text{-----(28)}$$

where the complex nonlinear coefficient $A = A_n + iA_p$ with $|A| = \sqrt{A_n^2 + A_p^2}$. Equation (28) describes the evolution of weakly nonlinear small-amplitude DASWs in an unmagnetized collisionless dusty plasma with a pair of trapped ions and the effect of dust charge variation. The nonlinear coefficient *A* becomes complex due to the vortex-like distributions of both positive and negative ions [25]. In absence of one of them, A becomes real and one can then obtain compressive or rarefactive solitary waves with positive or negative potential. A stationary solition solution of Eq. (28) can be obtained with its absolute value as [25]

$$|\phi| = \phi_m^2 \sec h^4\left[(\xi - u_0 t)/w\right] \quad \text{-----(29)}$$

where $u_0$ is a constant, $\phi_m = 15u_0/(8|A|)$, and $w = \sqrt{(16B/u_0)}$ are the amplitude and width of the soliton respectively.

**IV. Derivation of Burgers equation**

In this section we consider a different scaling of the stretched coordinates to take into account the dissipative effects due to the dust charge dynamics. Here, the independent variables $\xi$ and $\tau$ are stretched as

$$\xi = \varepsilon^{1/2}\left(I_x x + I_y y + I_z z - MT\right) \text{ and } \tau = \varepsilon T \quad \text{----- (30)}$$

The expansion for the dependent variables will remain the same as in Sec. IV. Thus, we obtain the same first order quantities [Eqs. (18) and (19)] as before. However, in the next higher order of $\varepsilon$ we obtain from the Poisson equation and the dust charge evolution equation the following

$$\left(\sigma\mu_n + \mu_p\right)\phi^{(2)} - b_n\mu_n(\sigma)^{\frac{3}{2}}\left(\phi^{(1)}\right)^{\frac{3}{2}} + b_p\mu_p\left(-\phi^{(2)}\right)^{\frac{3}{2}} - \alpha\left(N_d^{(2)} + Z_d^{(2)}\right) = 0 \quad \text{----- (31)}$$

$$\frac{\partial Z_d^{(1)}}{\partial \xi} - \alpha(1+\sigma)\phi^{(2)} - \alpha b_p\left(-\phi^{(1)}\right)^{3/2} + \alpha b_n(\sigma)^{\frac{3}{2}}\left(\phi^{(1)}\right)^{3/2} - \alpha\left(\beta_p + \beta_n\right)Z_d^{(2)} = 0 \quad \text{---- (32)}$$

From Eq. (32) we find that the dynamical evolution part appears corresponding to the first order (instead of the second order as in the case of KdV equation) perturbation of the dust charge state. So, it follows that in the evolution equation for the first order perturbations, the dynamical part of the dust charge number will contribute to the dissipative term in the evolution equation.

Eliminating $V_j^{(2)}$ and $N_d^{(2)}$ from Eqs. (31) - (32) with the help of Eqs. (18) and (19) we obtain the following complex Burgers equation for positively and negatively charged dust grains $(\alpha = \pm 1)$ as

$$\frac{\partial \phi^{(1)}}{\partial \tau} + A_p\left(-\phi^{(1)}\right)^{1/2}\frac{\partial \phi^{(1)}}{\partial \xi} + A_n\left(\phi^{(1)}\right)^{1/2}\frac{\partial \phi^{(1)}}{\partial \xi} - C\frac{\partial^2 \phi^{(1)}}{\partial \xi^2} = 0 \quad \text{------ (33)}$$

Equation (33) can be rewritten as

$$\frac{\partial \phi^{(1)}}{\partial \tau} + A\left(\phi^{(1)}\right)^{1/2}\frac{\partial \phi^{(1)}}{\partial \xi} - C\frac{\partial^2 \phi^{(1)}}{\partial \xi^2} = 0, \quad \text{------ (34)}$$

where the nonlinear coefficient $A$ is the same as for the KdV equation (28) and the dissipation coefficient is

$$C = \frac{\left(3M^2 - I_x^2 5\sigma_d\right)\left(1+\sigma_n\right)}{6M\left\{\left(\beta_p + \beta_n\right)\left(\sigma\mu_n + \mu_p\right) + \left(1+\sigma_n\right)\right\}} \quad \text{------ (35)}$$

Equation (34) governs the dynamical evolution of DA shocks in a dusty pair-ion plasma with dust charge fluctuations. The latter contribute to the dissipative term and so to the generation of shock waves. In order to obtain a stationary solution of Eq. (34) we use the transformation $\eta = \xi - u_0\tau$ and consider $\phi^{(1)}(\xi,\tau) = \psi(\chi)$ which gives

$$-C\frac{d\psi}{d\chi} + |A|\frac{2\psi^{\frac{3}{2}}}{3} + (-u_0)\psi = 0 \quad \text{------ (36)}$$

Next, we apply the well-known *tanh*-method in which we introduce $z = \tanh(\chi)$ and $\psi(\chi) = W(z)$, so that Eq. (36) reduces to

$$\frac{2|A|}{3}W^{\frac{3}{2}} - C(1-z^2)\frac{dW}{dZ} + (-u_0)W = 0 \qquad \text{------ (37)}$$

To find a series solution of Eq. (37) we substitute $W(z) = \sum_{r=0}^{\infty} a_r z^{\rho+r}$ in it. In the leading order analysis of finite terms, we have $r = 2$ and $\rho = 0$ and then $W(z)$ becomes $W(z) = a_0 + a_1 z + a_2 z^2$. Now substituting the expression $W(z) = a_0 + a_1 z + a_2 z^2 = a_0(1-z)^2$ in Eq. (37), we obtain the following stationary solution of the complex modified Burgers Eq. (34) as

$$|\phi| = \phi_n^2 \{1 - tanh(\chi/c)\}^2 \qquad \text{----- (38)}$$

where $\phi_n = 3u_0/4|A|$ and $c = 4C/u_0$ are the height and thickness of the shock wave solution respectively.

**V. Results and Discussion:**

The effects of various plasma parameters on the propagation characteristics of electrostatic solitary and shock waves are examined numerically on the basis of the stationary solutions (29) and (38). The charged dust particles in the electron free plasma are considered as uniform in size with fluctuating charge in presence of singly charged positive and negative ions. Figures 1(a) and 1(b) show the variations of the nonlinear co-efficient *A* with trapped positive (negative) ion parameter at different density ratios for positive and negative ions. In Fig. 1(a), the lines marked 1 and 2 [3 and 4] are for a constant $\sigma_p$ (the positive ion trapped parameter) [$\sigma_n$ (the negative ion trapped parameter)] and with the variation of $\sigma_n$ [$\sigma_p$] at different values of the number density ratio $\mu_p$. The parameter $\mu_n$ corresponding to the negative ion number density is treated as constant for each of these subplots. Similarly, subplot 1(b) shows the lines at different values of the number density ratio $\mu_n$ keeping $\mu_p$ as constant. From both the subplots we find that **the** decrease in the value of the nonlinear coefficient for a particular value of $\sigma_p$ with $\sigma_n$ reaching towards the value in thermal equilibrium is faster than its value for a particular value of $\sigma_n$ with $\sigma_p$ reaching towards the value in thermal equilibrium. This is due to the fact that as and when the system advances towards the thermal equilibrium, the nonlinearity of the system tends to get reduced. Such a decrement of *A* with $\sigma_n$ is also clear from the expression $A_n$ of *A* in which the contribution of the trapped negative ion appears through a reducing factor with $\sigma_n$. Thus, the role of trapped negative ion species is to reduce the nonlinearity in the KdV equation. Such a reduction is lower than that comes from the trapped positive ion species. Figure 2 (3) exhibits the variation of the solitary wave potential $\phi$ with $\sigma_p$ ($\sigma_n$) keeping the other fixed, *i.e.*, a fixed value of $\sigma_n$ ($\sigma_p$). These trapped parameters range from a value corresponding to hump shaped type with Boltzmann distribution to that giving rise the thermal equilibrium through a flat topped distribution. The other plasma parameters are considered as, $n_{d0} = 10^{10} \, m^{-3}$, $z_{d0} = (\pm)2 \times 10^4 e$, $T_i = 0.16 \, eV$, $T_d = 0.06 \, eV$ [6,7,26]. In both the cases, it is seen that as the system moves from a dip shaped distribution to Boltzmann distribution i.e. towards the thermal equilibrium, the solitary wave potential shows a significant increase in its absolute value. This is because, with the system advancing towards the thermal equilibrium, the nonlinearity decreases, which gives rise an increment of the localized wave potential of the solitary wave. However interestingly, the solitary solution shows a higher (lower) height and lower (higher) width with changing the parameters corresponding to the positive (negative) trapped ions at a particular value of the other one.  This is also in accordance with the physical features discussed in the variations of the nonlinear coefficient *A* demonstrated in Figures 1(a) and 1(b). As the nonlinear coefficient *A* appears in the solitary wave solution as an inverse function, so we find a greater height of the solitary wave with negative ion trapped parameter (Fig. 2) than compared to the positive ion trapped parameter (Fig. 3).

Figures 4 and 5 show the profiles of the absolute values of the solitary wave potential for different values of $\sigma_p$ and $\sigma_n$. It is seen that both the amplitude and the width of solitons are changed significantly due to the change of parameters corresponding hump type or flat topped distributions of ions.  From Figs. 4 and 5, we find that as and when the nonlinearity of the system increases (decreases), the solitary wave potential (absolute value) decreases (increases) and this increase (decrease) of the nonlinearity depends on the consideration of the dip-shaped (Boltzmanian) distribution of the trapped plasma species.

Figures 6 to 9 exhibit the profiles of the DA shocks given by Eq. (38).   It is seen from Figs. 6 and 8 that the height of the shock front gets reduced compared to the solitary wave potentials.  This is due to the dissipative factor which comes into play in the formation of shock waves primarily. A similar (to the solitary wave) trend of increase or decrease in the wave potential is also seen. For example, if we consider Figs. 6 and 8, it is seen that as we move from a dip shaped distribution of positive ions (negative ions) towards thermal equilibrium at a particular value of the trapped negative (positive) ion parameter, the shock potential (absolute value) increases. The value of the shock potential becomes higher at a particular value of the trapped negative ion parameter than compared to that of the trapped positive ion parameter. The reason may be similar (except the magnitude) as discussed in case of Figs 2 and 3. Similarly, from Figs 7 and 9, it is seen that with the effects of ions having a flat topped distribution, the shock wave potential has the higher value (Fig. 7) than compared to the case of ions having a hump and/or Boltzmann distribution (Fig. 9). On the other hand, by comparing the profiles in Figs. 7 and 9, we see that in Fig. 7, the amplitude of the shock wave potential is higher for a flat topped distribution of positive ions and Boltzmann distribution of negative ions, while it is lower for flat topped distribution of positive ions and a dip shaped distribution of negative ions. However, Fig. 9, it can be concluded that a dip-shaped distribution of negative ions and Boltzmann distribution of positive ions result in an enhancement of the potential compared to that due to a dip shaped distribution of positive ions. Similar to the solitary waves, and relying on the profiles in Figs. 7 and 9, it is seen that as and when the nonlinearity of the plasma system increases (decreases), the shock wave potential decreases (increases) and this increase (decrease) of the nonlinearity depends on the dip shaped (Boltzmanian) distribution of the trapped plasma species.

## VII. Conclusion:

We have investigated the weakly nonlinear propagation of small amplitude dust-acoustic solitary waves and shocks in a dusty pair-ion plasma with dust charge fluctuations due to trapped positive and negative ion species. Using the reductive perturbation technique, we have derived the evolution equations for these solitary waves and shocks in the forms of KdV and Burgers equations with a complex nonlinear coefficient. The latter becomes complex due to vortex-like distributions of two oppositely charged ion species. Travelling wave solutions of these KdV and Burgers equations are obtained and analysed numerically. The effects of the trapped positive (negative) ions with a hump (dip) shaped distributions as well as flat-topped and Boltzmannian ones on the profiles of the height and width of the solitary waves and shocks are also investigated. It is found that for a particular value of the positive (negative) trapped ion parameter with trapped negative (positive) ion distributions moving towards the thermal equilibrium, the nonlinear coefficient decreases sharply irrespective of any change in the positive (negative) ion density ratio. It is also seen that the change of distribution of the trapped positive (negative) ions from a hump (dip) shaped through flat-topped to Boltzmannian has weighty effect on the height and width of both solitary and shock wave potentials. The results may be useful for understanding the propagation characteristics of nonlinear dust-acoustic waves and shocks in laboratory plasmas where, e.g., $n_{p0} \sim 10^{14} m^{-3}$, $n_{n0} \sim 10^{14} m^{-3}$, $n_{d0} \sim 10^{10} m^{-3}$, $z_{d0} \sim 10^4 e$, $m_p/m_n = 28/300 \approx 0.4$, $T_p \sim 10^3 K$, $T_p \sim 3 \times 10^2 K$, and space plasmas, *e.g.*, in a dusty region at an altitude of about 95 km in the Earth's mesosphere [24,26,27] where $T_p \sim T_n \sim 200$ K, $m_n/m_p = 300/28 \approx 10.7$, $n_{n0} \sim 2 \times 10^{10}$ m$^{-3}$, $n_{p0} \sim 10^{10}$ m$^{-3}$, $z_d n_{d0} \sim 10^{10}$ m$^{-3}$.

**Acknowledgement:**
APM acknowledges support from UGC-SAP (DRS, Phase III) with Sanction order No. F.510/3/DRS-III/2015 (SAPI),   and UGC-MRP with F. No. 43-539/2014 (SR) and FD Diary No. 3668.

**Figure Captions:**

FIG.1. Plot of the Mach number against $\sigma$ for different values of $\mu_n$. For dotted line (black) $\mu_n$ = 0.03, solid line (blue) $\mu_n$ = 0.13, dash-dotted line (red) $\mu_n$ = 0.3 and dashed line (green) $\mu_n$ = 1.3.

FIG.2(a). For plot (1) and (2), $\sigma_p$ is 0.5, so X-axis variation is with $\sigma_n$ with $\mu_p$ is 0.6 and1 respectively. Similarly for plot (3) and (4), $\sigma_n$ is 0.5, so X-axis variation is with $\sigma_p$ with $\mu_p$ is 0.6 and1 respectively.

FIG.2(b). For plot (1) and (2), $\sigma_p$ is 0.5, so X-axis variation is with $\sigma_n$ with $\mu_n$ is 0.6 and1 respectively. Similarly for plot (3) and (4), $\sigma_n$ is 0.5, so X-axis variation is with $\sigma_p$ with $\mu_n$ is 0.6 and1 respectively.

FIG.3. Variation of solitary wave potential with $\sigma_p$ and $\chi$ at specific value of $\sigma_n$.

FIG.4. Variation of solitary wave potential with $\sigma_n$ and $\chi$ at specific value of $\sigma_p$.

FIG.5. Variation of solitary wave potential with $\chi$ at different values of $\sigma_p$ and $\sigma_n$ . Black $\sigma_p = 0$ & $\sigma_n = 0.2$, Dashed $\sigma_n = 0$ & $\sigma_p = 0.4$ , Blue $\sigma_n = 0$ & $\sigma_p = 0$, Dashed dot $\sigma_n = 0$ & $\sigma_p = -0.4$ , Dot $\sigma_n = -0.2$ & $\sigma_p = 0$

FIG.6. Variation of solitary wave potential with $\chi$ at different values of $\sigma_p$ and $\sigma_n$. Blue $\sigma_p = -0.2$ & $\sigma_n = 0.5$, Dashed $\sigma_p = -0.2$ & $\sigma_n = -0.2$, Black $\sigma_p = -0.2$ & $\sigma_n = -0.5$ Dashed dot $\sigma_p = -0.5$ & $\sigma_n = -0.2$, Dotted $\sigma_p = -0.5$ & $\sigma_n = 0.2$.

FIG.7. Variation of shock wave potential with $\chi$ and $\sigma_p$ at a specific value of $\sigma_n$.

FIG.8. Variation of shock wave potential with $\chi$ at different values of $\sigma_p$ and $\sigma_n$. Dashed $\sigma_p = 0$ & $\sigma_n = 0.4$, Black $\sigma_p = 0.4$ & $\sigma_n = 0$, Dotted $\sigma_p = 0$ & $\sigma_n = 0$, Dashed dot $\sigma_p = -0.4$ & $\sigma_n = 0$, Blue $\sigma_p = 0$ & $\sigma_n = -0.2$

FIG.9. Variation of shock wave potential with $\chi$ and $\sigma_n$ at a specific value of $\sigma_p$.

FIG.10. Variation of shock wave potential with $\chi$ at different values of $\sigma_p$ and $\sigma_n$. Dashed-dot $\sigma_n = -0.2$ & $\sigma_p = 0.5$, Black $\sigma_n = -0.2$ & $\sigma_p = -0.2$, Dashed $\sigma_n = -0.2$ & $\sigma_p = -0.5$, Blue $\sigma_n = -0.5$ & $\sigma_p = 0.2$, Dotted $\sigma_n = -0.5$ & $\sigma_p = -0.2$.

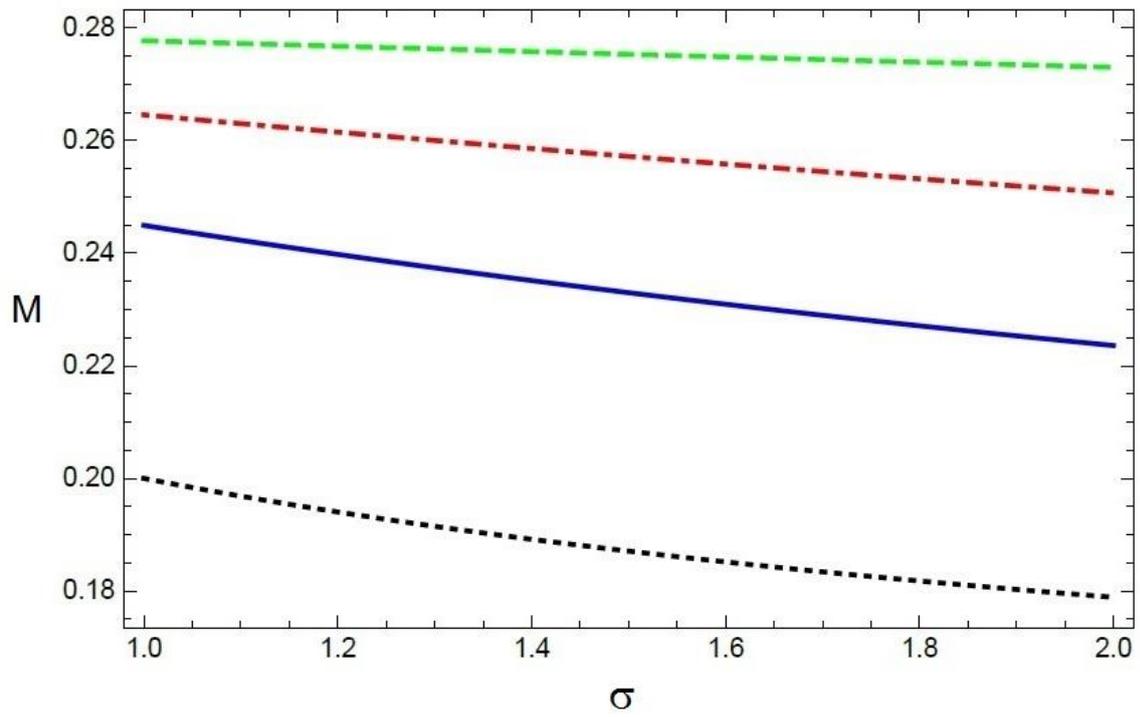

FIG.1. Plot of the Mach number against $\sigma$ for different values of $\mu_n$. For dotted line (black) $\mu_n$ = 0.03, solid line (blue) $\mu_n$ = 0.13, dash-dotted line (red) $\mu_n$ = 0.3 and dashed line (green) $\mu_n$ = 1.3.

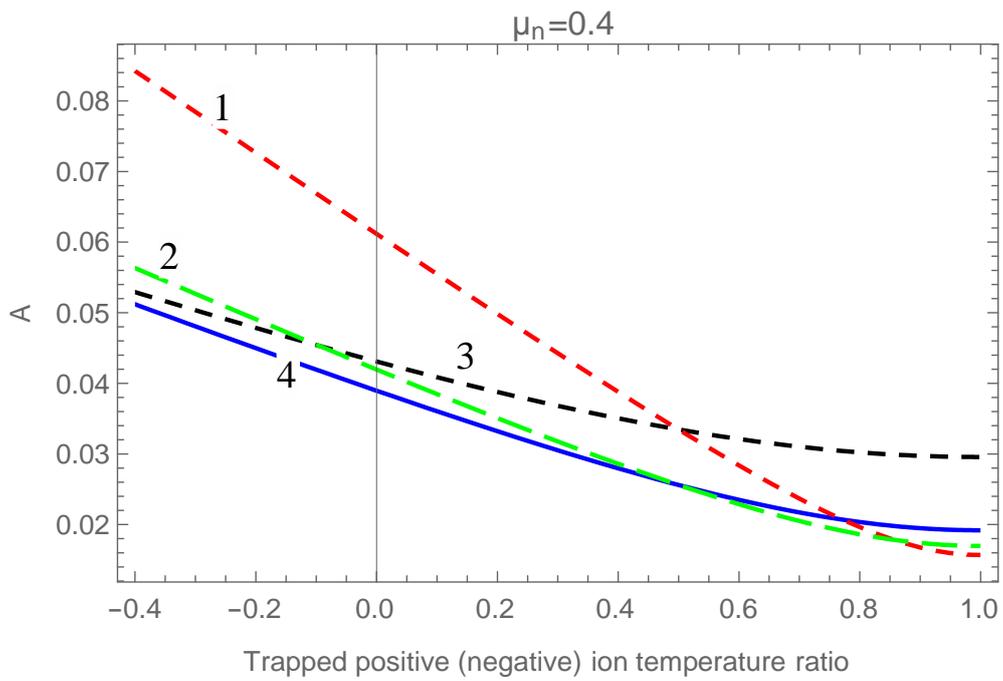

FIG.2(a). For plot (1) and (2), $\sigma_p$ is 0.5, so X-axis variation is with $\sigma_n$ with $\mu_p$ is 0.6 and 1 respectively. Similarly for plot (3) and (4), $\sigma_n$ is 0.5, so X-axis variation is with $\sigma_p$ with $\mu_p$ is 0.6 and 1 respectively.

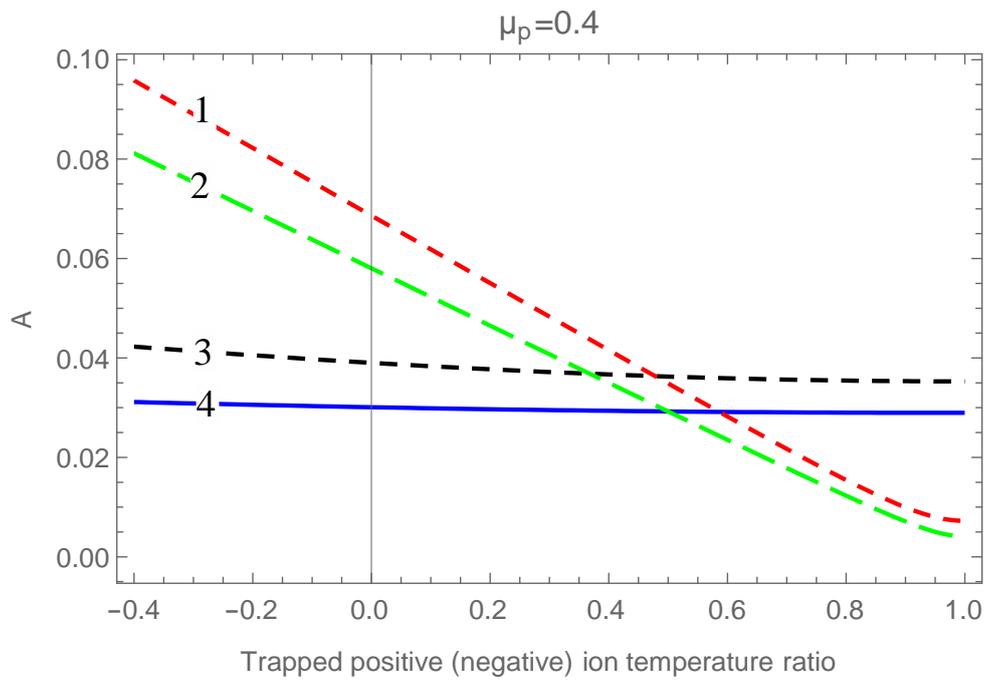

FIG.2(b). For plot (1) and (2), $\sigma_p$ is 0.5, so X-axis variation is with $\sigma_n$ with $\mu_n$ is 0.6 and 1 respectively. Similarly for plot (3) and (4), $\sigma_n$ is 0.5, so X-axis variation is with $\sigma_p$ with $\mu_n$ is 0.6 and 1 respectively.

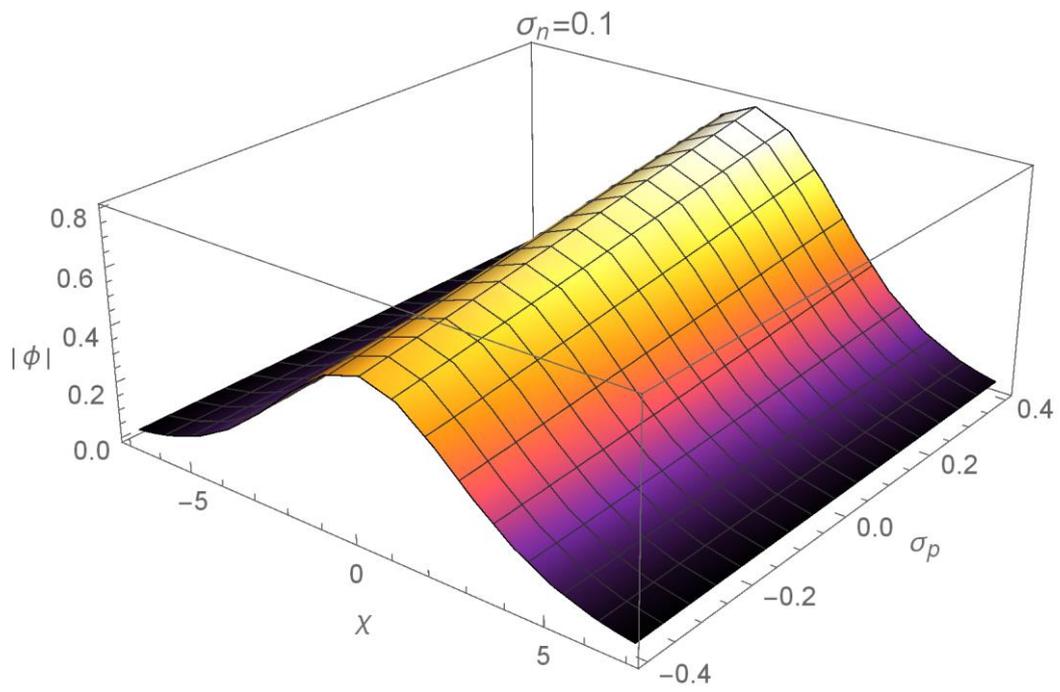

FIG.3. Variation of solitary wave potential with $\sigma_p$ and $\chi$ at specific value of $\sigma_n$.

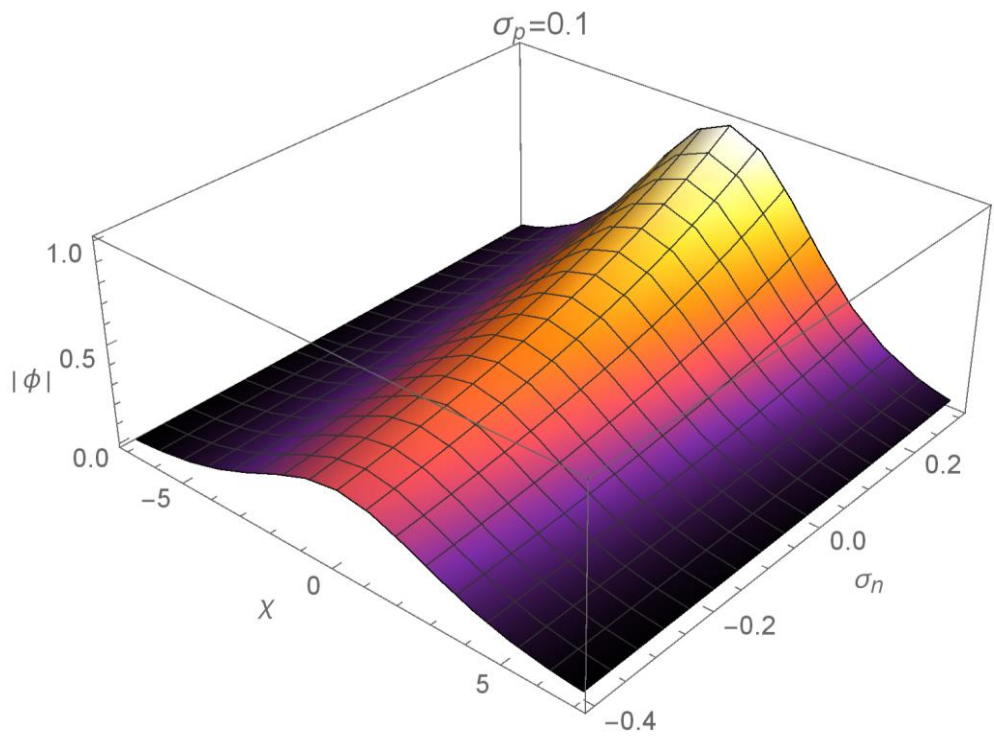

FIG.4. Variation of solitary wave potential with $σ_n$ and $χ$ at specific value of $σ_p$.

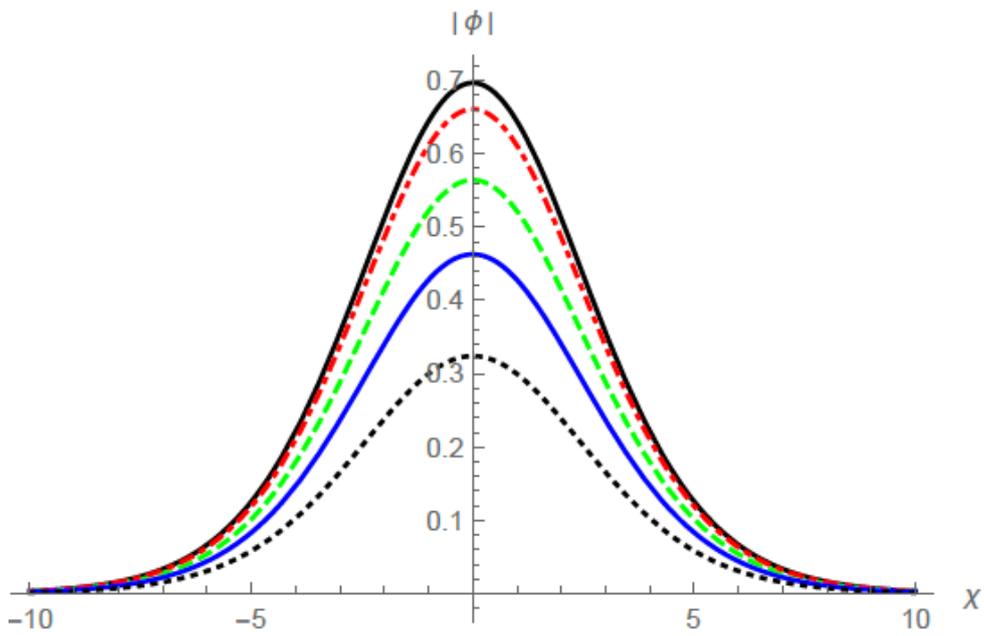

FIG.5. Variation of solitary wave potential with χ at different values of $\sigma_p$ and $\sigma_n$. Black $\sigma_p = 0$ & $\sigma_n = 0.2$, Dashed $\sigma_n = 0$ & $\sigma_p = 0.4$, Blue $\sigma_n = 0$ & $\sigma_p = 0$, Dashed dot $\sigma_n = 0$ & $\sigma_p = -0.4$, Dot $\sigma_n = -0.2$ & $\sigma_p = 0$

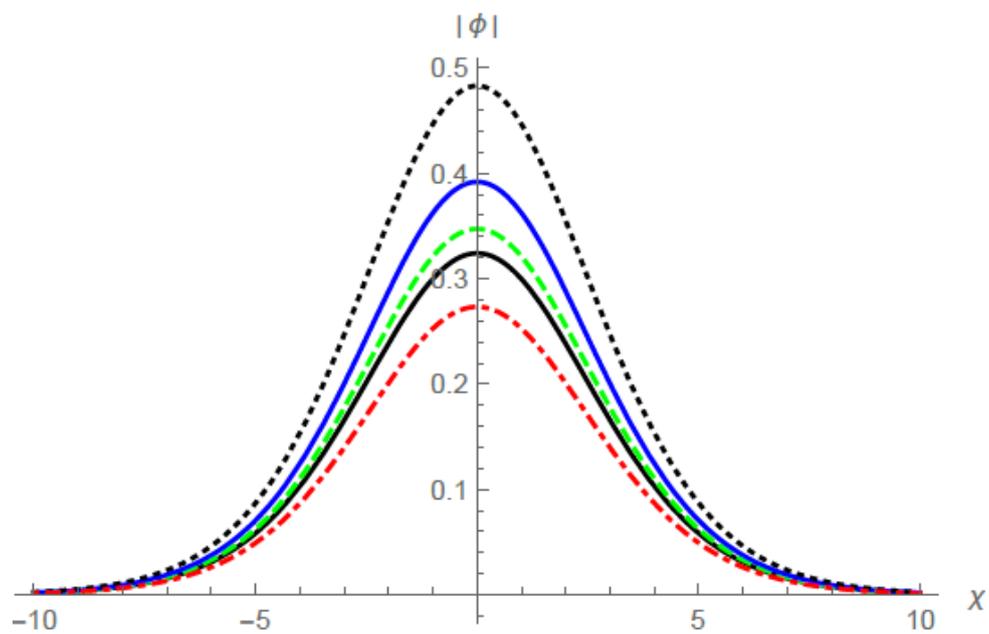

FIG.6. Variation of solitary wave potential with χ at different values of $\sigma_p$ and $\sigma_n$. Blue $\sigma_p = -0.2$ & $\sigma_n = 0.5$, Dashed $\sigma_p = -0.2$ & $\sigma_n = -0.2$, Black $\sigma_p = -0.2$ & $\sigma_n = -0.5$ Dashed dot $\sigma_p = -0.5$ & $\sigma_n = -0.2$, Dotted $\sigma_p = -0.5$ & $\sigma_n = 0.2$.

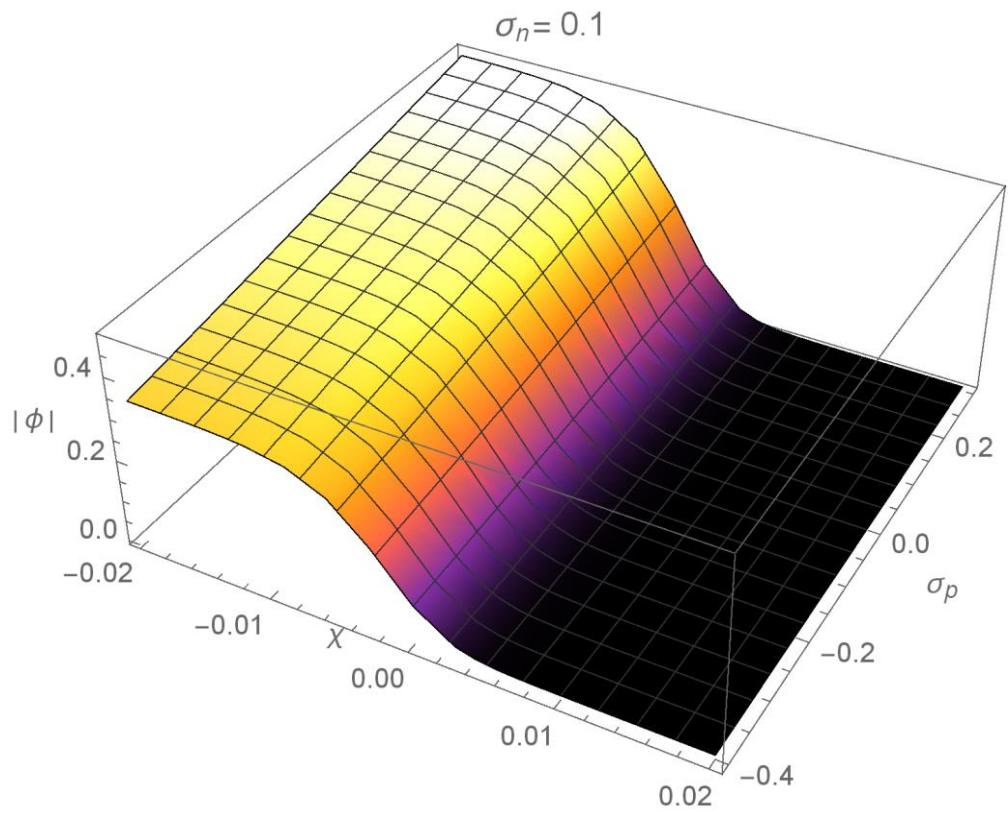

FIG.7. Variation of shock wave potential with $\chi$ and $\sigma_p$ at a specific value of $\sigma_n$.

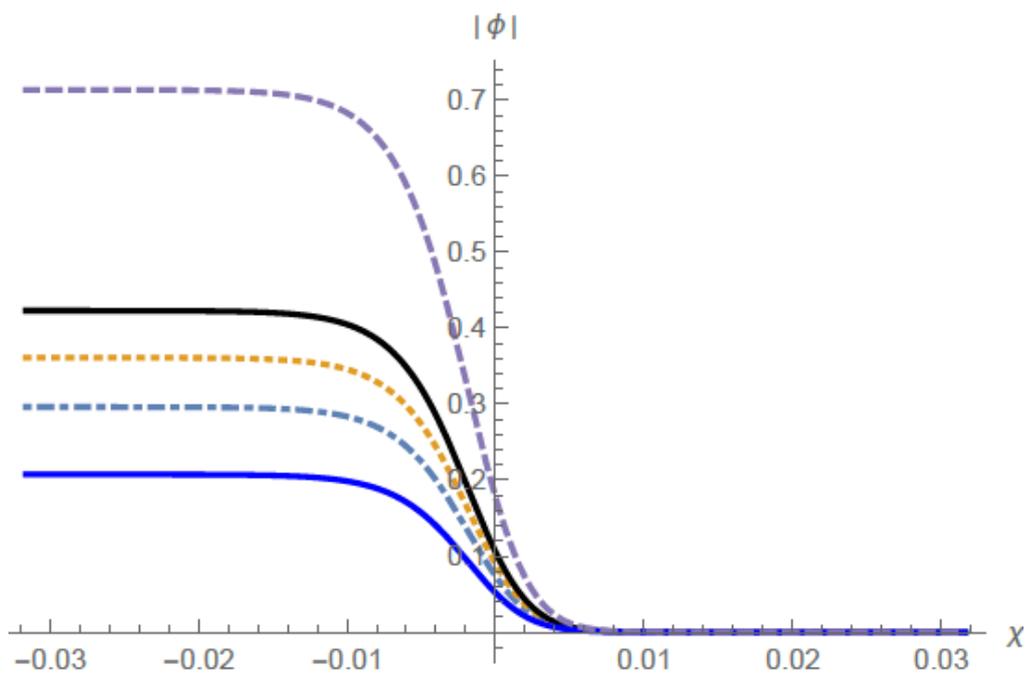

FIG.8. Variation of shock wave potential with χ at different values of $\sigma_p$ and $\sigma_n$. Dashed $\sigma_p = 0$ & $\sigma_n = 0.4$, Black $\sigma_p = 0.4$ & $\sigma_n = 0$, Dotted $\sigma_p = 0$ & $\sigma_n = 0$, Dashed dot $\sigma_p = -0.4$ & $\sigma_n = 0$, Blue $\sigma_p = 0$ & $\sigma_n = -0.2$

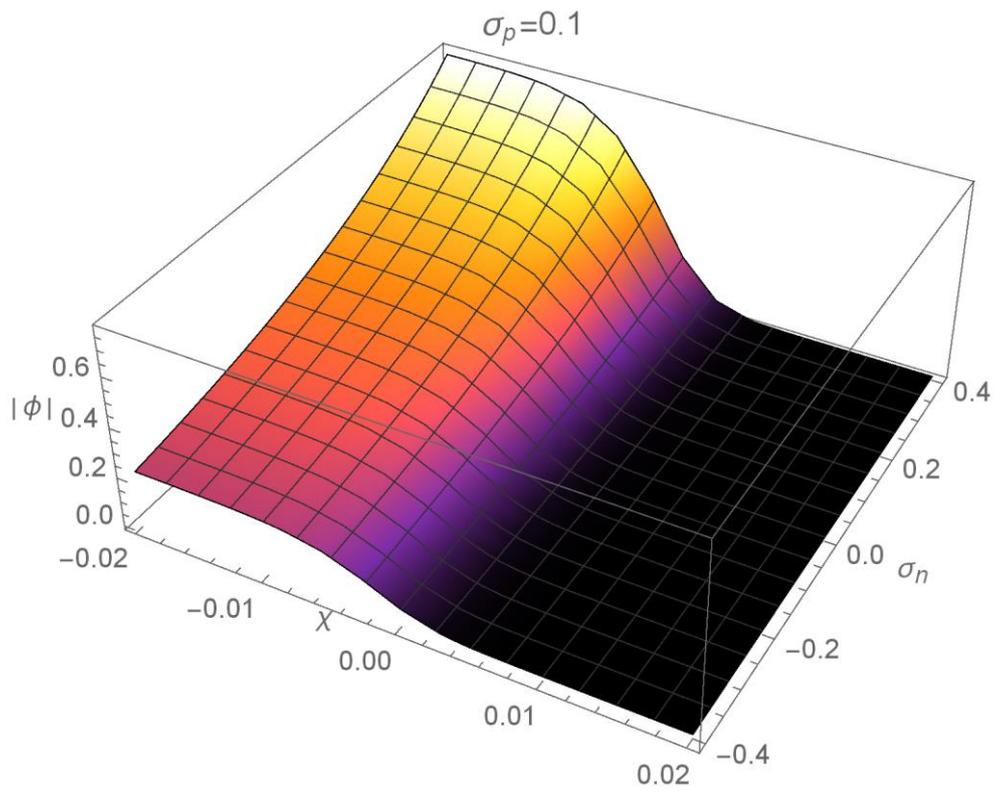

FIG.9. Variation of shock wave potential with $\chi$ and $\sigma_n$ at a specific value of $\sigma_p$.

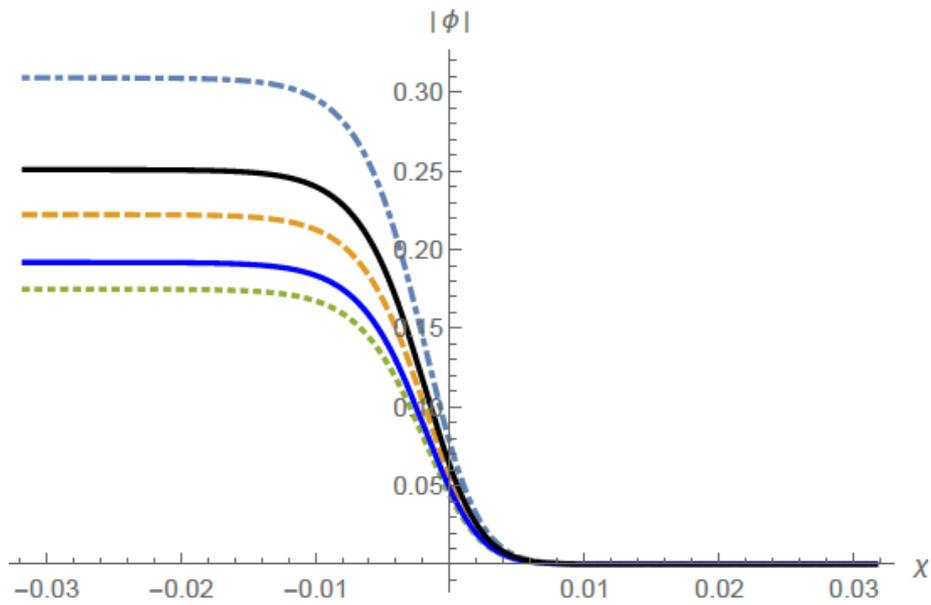

FIG.10. Variation of shock wave potential with χ at different values of $σ_p$ and $σ_n$. Dashed-dot $σ_n = -0.2$ & $σ_p = 0.5$, Black $σ_n = -0.2$ & $σ_p = -0.2$, Dashed $σ_n = -0.2$ & $σ_p = -0.5$, Blue $σ_n = -0.5$ & $σ_p = 0.2$, Dotted $σ_n = -0.5$ & $σ_p = -0.2$.